# Energy-efficient Software-defined 5G/6G Multimedia IoV: PID controller-based approach

Ahmadreza Montazerolghaem

*Abstract*—The rapid proliferation of multimedia applications in smart city environments and the Internet of Vehicles (IoV) presents significant challenges for existing network infrastructures, particularly with the advent of 5G and emerging 6G technologies. Traditional architectures struggle to meet the demands for scalability, adaptability, and energy efficiency required by data-intensive multimedia services. To address these challenges, this study proposes an innovative, energy-efficient framework for multimedia resource management in software-defined 5G/6G IoV networks, leveraging a Proportional-Integral-Derivative (PID) controller. The framework integrates Software-Defined Networking (SDN) and Network Functions Virtualization (NFV) technologies to enable centralized and adaptive control over network resources. By employing a PID controller, it dynamically manages load distribution and temperature, ensuring balanced resource allocation and minimizing energy waste. Comprehensive simulations validate the framework's effectiveness, demonstrating significant improvements in load balancing, CPU utilization, and energy consumption compared to traditional methods. For instance, under heavy traffic conditions, the proposed framework maintained resource efficiency, reducing power consumption by up to 30% and achieving nearly equal load distribution across all network components. Additionally, the controller exhibited exceptional scalability, effectively responding to over 98% of vehicle requests even in scenarios of extreme traffic demand.

## I. INTRODUCTION

**M**ULTIMEDIA applications, particularly in smart city environments and the IoV, have rapidly increased the demand for robust, flexible, and efficient network infrastructures in recent years [1]. With the advent of 5G and on the horizon 6G technologies, vehicular networks will be able to evolve into dynamic, adaptive ecosystems that deliver seamless, high-quality multimedia experiences [2]. 5G technology has introduced ultra-reliable low-latency communication (URLLC) and enhanced mobile broadband (eMBB), both of which are essential for IoV systems that rely on real-time, data-intensive multimedia services [3]. The anticipated 6G technology will push these capabilities further, enabling extreme data rates and pervasive connectivity, and thereby positioning itself as a backbone for future multimedia-driven IoV networks [4]. However, traditional network architectures lack the scalability and adaptability needed to meet these evolving demands, particularly as multimedia applications place unprecedented strain on bandwidth, latency, and energy resources [5].

A. Montazerolghaem is with the Faculty of Computer Engineering, University of Isfahan, Isfahan, Iran. E-mail: a.montazerolghaem@comp.ui.ac.ir

To address these challenges, SDN and NFV have emerged as transformative frameworks [6]. SDN enables centralized, programmable control over network resources, allowing real-time adaptation to fluctuating traffic patterns and demand peaks, while NFV decouples network functions from dedicated hardware, enabling flexible deployment and scaling of services as virtualized network functions (VNFs) [7]. Together, SDN and NFV create an energy-efficient infrastructure ideal for managing the complex, data-heavy demands of multimedia traffic in IoV environments. Yet, as the volume and complexity of multimedia applications continue to grow, existing SDN/NFV frameworks face critical challenges in balancing resource allocation with the energy efficiency required in smart cities. As 5G and 6G networks mature, there lies a transformative opportunity to redefine multimedia resource management, ensuring high quality of experience (QoE) for users while meeting sustainability goals through optimized resource and energy management [8]. The demand for innovative solutions that can dynamically manage multimedia traffic and resource allocation, leveraging the advancements of SDN, NFV, and next-generation networks, has thus become an urgent priority in the design of IoV architectures. Furthermore, the lack of intelligent control mechanisms, such as PID controllers, limits the ability to proactively manage load distribution and energy consumption, exacerbating issues of inefficiency and degradation of service quality.

This paper presents an innovative framework for managing multimedia resources in software-defined 5G/6G IoV networks, utilizing an adaptive PID controller. Unlike conventional, static methods, this proactive approach dynamically optimizes load distribution and energy use in real-time. Integrating SDN and NFV capabilities, the framework enhances operational efficiency while reducing energy waste. A key feature is its dual focus on adaptive load and temperature management, using predictive analytics to respond effectively to changing multimedia demands and improve stability and resource use.

### A. Motivations

The rapid evolution of multimedia content and its integration into everyday life, particularly within the context of the Internet of Things (IoT), has created unprecedented demands on network infrastructure [9]. As vehicles become smarter and more connected, the demand for real-time, high-quality multimedia streaming and communication, such as augmented reality (AR)/virtual reality (VR)-enhanced navigation for advanced driver-assistance systems (ADAS), 4K/8K immersive infotainment via vehicle-to-vehicle (V2V) sharing, and ultra-low-latency video feeds for collision avoidance and emergency management, has surged [10], [11]. With the advent of 5G and the impending rollout of 6G technologies, there is a significant opportunity to enhance multimedia experiences for users in smart cities and interconnected environments [12]. This growth is driven by the proliferation of multimedia applications, from streaming services to real-time interactive experiences, requiring robust and efficient resource management to meet quality of experience expectations.

As multimedia traffic continues to surge, traditional network architectures face challenges in scalability, flexibility, and energy efficiency [13]. SDN and NFV emerge as transformative solutions that enable dynamic resource allocation and optimized management of network functions in the external data network, where multimedia flows originate and terminate, with 5G/6G cellular networks providing connectivity. By leveraging SDN's centralized control and NFV's capability to virtualize network services, we can create adaptable frameworks that respond to the real-time demands of multimedia applications in the Internet of multimedia vehicles. Moreover, the integration of intelligent control mechanisms, such as PID controllers, facilitates the proactive management of load distribution and energy consumption. This approach not only enhances operational efficiency but also significantly reduces energy waste, aligning with global sustainability goals. The motivation for this research lies in addressing the intricate challenges posed by the intersection of multimedia applications, IoV, and advanced cellular networks. Our proposed framework (Fig. 1) aims to establish a resilient and energy-efficient multimedia resource management system, thus paving the way for the future of connected vehicular networks and improving user experiences in a rapidly evolving digital landscape.

### B. Contributions

This paper presents a novel, energy-efficient framework for multimedia resource management in software-defined 5G/6G IoV networks, utilizing an adaptive PID controller-based approach. The major contributions of this work are as follows:

*1) Development of an Energy-Optimized 5G/6G Multimedia IoV Framework:* we propose a modular, SDN-based architecture for large-scale multimedia IoV systems, addressing real-time demands due to vehicular mobility. This architecture incorporates load balancing and energy management to enhance the quality of experience for vehicular users. By leveraging SDN's centralized control capabilities and NFV, this predictive framework dynamically allocates resources, ensuring balanced server loads and improved energy efficiency across network components.

*2) Integration of a PID Controller for Intelligent Load and VNF Optimization:* the framework introduces a modular PID controller mechanism that autonomously adjusts load distribution and VNF placement based on real-time monitoring and predictions. The PID controller uses system temperature as a metric to optimize resource allocation and energy efficiency, reducing waste during low demand and preventing overload during high demand. This approach enhances the operational stability and efficiency of multimedia services within the IoV environment.

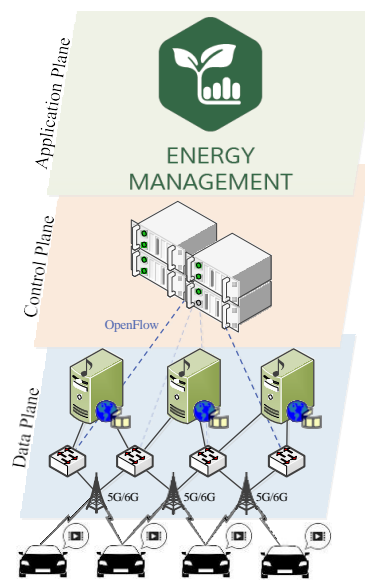

Figure 1. Proposed framework.

*3) Performance Validation in a Simulated IoV Environment:* extensive simulations validate the proposed framework's effectiveness in minimizing energy consumption and improving server response times compared to traditional load management methods. The results demonstrate the feasibility of this PID-controlled approach in real-world IoV scenarios, making it a promising solution for next-generation energy-efficient 5G/6G multimedia networks.

### C. Organization

The literature review presented in Section II examines the relevant research background and context for this study. Section III outlines the exquisite modular energy-efficient framework and controller design for a software-defined 5G/6G multimedia IoV network. This section introduces new algorithms proposed for each of the system's modules. Section IV deploys the proposed framework, assesses the resulting performance metrics, and conducts a detailed evaluation. Finally, Section V provides conclusions and identifies potential directions for future research.

## II. RELATED WORK

The integration of 5G/6G technologies into vehicular networks marks a significant progression in intelligent transportation systems. This review synthesizes current research, emphasizing the interplay of these next-generation technologies with IoV and identifying existing gaps.

### A. SDN and NFV in Intelligent Transportation

[14] presents a comprehensive design and prototyping of a Software Defined Vehicular Network (SDVN), emphasizing the integration of SDN principles to enhance vehicle communication through real-time routing and mobility management. It evaluates the performance of this architecture using real hardware, demonstrating its potential to optimize data transmission and improve connectivity in vehicular environments. [15] introduces the Reliability-Aware Flow Distribution Algorithm (RAFDA) for SDN-enabled Fog Computing in smart cities, optimizing IoT traffic routing by considering link reliability and

additional constraints like traffic load and bandwidth, thereby enhancing network performance. Bo Li et al. [16] present a novel methodology for dynamic controller placement in the Software-Defined Internet of Vehicles (SD-IoV). They utilize Mobile Edge Computing (MEC) and a multi-agent deep Q-learning network (MADQN) to optimize delay, load balancing, and reliability, achieving superior performance compared to existing baseline strategies. [17] proposes ARTNet, an AI-based framework for resource allocation and task offloading in Software-Defined Vehicular Fog Computing (SDV-F) networks. ARTNet leverages reinforcement learning to optimize load balancing and minimize end-to-end delay in dynamic IoV environments.

*B. 5G and 6G Technologies in IoV*

The study by [18] introduces an AI-driven 6G network management system employing Federated Learning to facilitate autonomous resource optimization. This distributed approach enhances energy efficiency and minimizes operational costs, especially in bandwidth-intensive applications such as virtual reality (VR) streaming. Early experiments show promise, though scalability and privacy concerns need more exploration in real-world settings. The article by [19] introduces a DRL-driven framework designed to enhance low-latency content delivery for 6G vehicular IoT networks. By dynamically selecting between vehicle-to-vehicle, infrastructure, and network modes, the approach optimizes data throughput and minimizes delay, outperforming conventional methods in simulated tests. However, it relies solely on simulation, and potential challenges with scalability and latency may arise in dense, high-traffic environments. [20] introduces FLOR, a Federated Q-Learning-based framework for efficient computation offloading and resource management in 6G vehicular networks. FLOR optimizes resource use and reduces latency across V2X communications by estimating upper delay bounds and allocating resources adaptively. While simulation results show improved performance over existing methods, real-world scalability and potential latency challenges under high network load remain areas for further investigation. To optimize task migration in the IoV, [21] presents a Network in Box (NIB) framework that integrates 6G technology with edge computing. It utilizes the Pareto envelope-based selection algorithm (PESA-II) for efficient resource allocation, validated with Shanghai datasets. However, its findings are limited by the use of a single dataset and insufficient analysis of scalability and comparison with other methods. The work in [22] proposes a UAV-assisted content delivery system for 5G networks, employing Knapsack optimization and Zipf distribution for efficient caching. The approach leverages a Stackelberg game-theoretic model to optimize storage allocation while integrating D2D communication to improve local content accessibility and overall network efficiency.

*C. Multimedia Resource Management in IoV*

In [23], a smart vehicular algorithm (SVA-IoT) is presented to optimize IoT-driven autonomous vehicle transportation systems using multimedia sensors. It focuses on improving resource management, enhancing communication reliability, and ensuring data security through digital certificates. The paper's validation relies on simulations, potentially limiting real-world applicability, and may not address all challenges in dynamic traffic environments. The goal of the proposed architecture in [24] is to provide high-quality multimedia streaming while optimizing energy usage within the IoV network. It achieves this by dynamically provisioning resources in response to traffic changes, ensuring efficient resource allocation and management. However, it may not fully address the limitations of existing non-SDN-based methods, which can struggle with real-time network state updates and orchestration. The study by [25] introduces a traffic offloading mechanism optimized for video quality in 5G networks, leveraging dual connectivity across macro and small cells. The system incorporates fountain coding to address packet loss and implements SDN-based resource management, showing superior performance relative to conventional offloading approaches.

*D. Energy Efficiency in IoV*

Chinmay Chakraborty and colleagues [26] investigate a hybrid optimization approach combining particle swarm optimization and genetic algorithms to improve energy efficiency in edge server clusters. Their proposed method aims to reduce operational costs while ensuring timely task execution and balanced workload distribution. The study focuses on the transfer of IoV tasks from cloud servers to edge servers via cellular networks, considering the challenges of network latency and limited processing capabilities of edge devices. [27] explores optimization techniques for resource allocation and mode selection in heterogeneous V2X networks to boost spectral and energy efficiency. The work presents a Deep Reinforcement Learning (DRL) framework that dynamically coordinates interference between V2I and V2V communications, balancing V2I throughput demands with V2V quality-of-service requirements. The developed AO-DDQN algorithm allows Vehicle User Equipment (VUEs) to make intelligent decisions by processing both real-time network parameters and historical performance data. [28] explores the potential of 6G networks to enhance IoT performance and security through energy harvesting and mobile edge computing. The proposed framework aims to optimize resource allocation and security measures. However, the reliance on accurate energy predictions, the adaptability of security configurations, and the complexity of implementation pose significant challenges. Additionally, the scope of security considerations and the trade-off between security and quality of service (QoS) require further investigation. The simulation-based evaluation may not fully capture real-world complexities.

*E. Outcomes*

In summary, while substantial progress has been made in the integration of SDN and NFV within IoV frameworks, there remains a critical need for research that focuses on energy efficiency, real-time resource management, and empirical validation of proposed models. Addressing these gaps will be essential for developing robust, adaptive multimedia resource management systems in next-generation vehicular networks.

III. PROPOSED FRAMEWORK

The network discussed in this paper is a large-scale network of multimedia vehicles with cellular communication for sending and receiving OTT (over-the-top) multimedia traffic from multimedia servers through a data network of OpenFlow switches and 5G/6G cellular network infrastructure. The primary goal of this paper is to enhance the QoE for vehicular users by managing multimedia server resources. In this context, both server load balancing and energy consumption reduction are considered. The proposed framework utilizes SDN technology. This enables a centralized global view of the data network resources for multimedia traffic. Consequently, it facilitates unified resource management and traffic distribution. The Fig. 1 illustrates an overview of the proposed framework, which comprises three layers. Data (Infrastructure) layer includes multimedia servers and multimedia vehicles interconnected by OpenFlow switches and cellular network equipment[1]. In SDN, traditional switches are replaced by OpenFlow switches. These switches are programmed using the standardized and widely used OpenFlow protocol, as well as an SDN controller. The OpenFlow protocol comprises a set of messages, such as *Features-request/reply*, *Packet-In*, and *Flow-mod*. These messages primarily serve to query switches or install instructions on them. The SDN controller, residing at the control plane, is responsible for collecting control data from the data network infrastructure. This data includes the amount of multimedia requests, the amount of resources available to servers, switch topology, link capacity, and more. The SDN controller gathers this data through OpenFlow protocol messages. In addition to data collection, the SDN controller also controls the switches. In this context, the necessary control decisions and instructions are made by application-level services based on the collected data. The application plane has comprehensive control over all multimedia data network information and serves as the system's mastermind. Consequently, intelligent algorithms are executed at this level. These algorithms (applications/modules) include routing, traffic prediction, load balancing, energy management, and more.

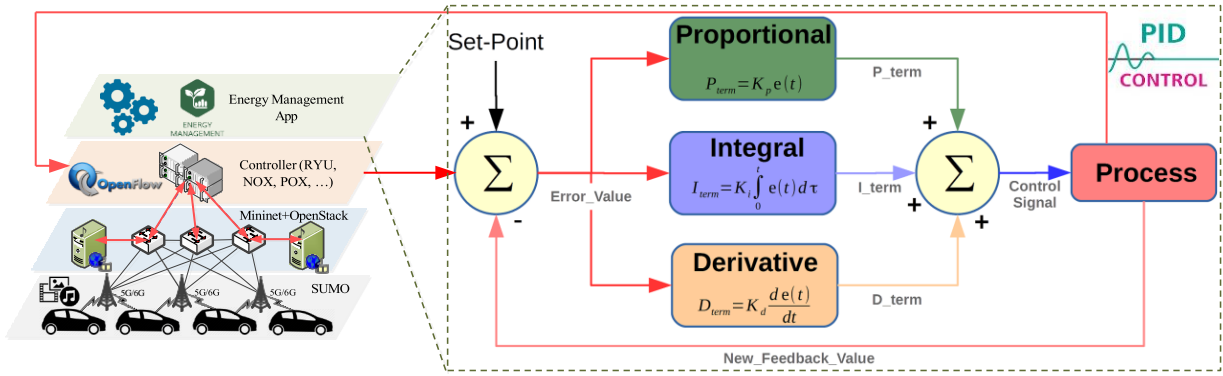

Figure 2. PID controller-based structure for management of proposed framework.

Fig. 2 shows a diagram illustrating the components and structure of a PID energy control system for a 5G/6G Multimedia IoV network, a novel feedback control mechanism. The diagram consists of several interconnected modules, each representing a specific function or operation within the PID control system. The central element is the "Process," which represents the system being controlled. Here system means the IoV infrastructure layer which includes OpenFlow switches and servers. The "PID Control" module is responsible for generating the appropriate control signal based on the error between the desired set-point and the actual process output. The input and output of this module are from/to the infrastructure layer, the details of which are explained in the following sections. Control commands are generated by the controller and sent to the infrastructure layer for execution. The "Proportional" component of the PID control adjusts the control signal in proportion to the current error, providing a rapid response to changes in the system. The "Integral" component accumulates the error over time, allowing the system to eliminate steady-state errors and achieve the desired set-point. The "Derivative" component anticipates future errors by monitoring the rate of change in the error, providing a damping effect and improving the system's stability. The diagram also includes additional elements, such as the "Set-Point" input, which represents the desired target value for the process, and the "Feedback" signal, which represents the actual output of the process. The "Error Value" is calculated by subtracting the Feedback signal from the Set-Point, and this error value is then used as the input to the PID control mechanism. The overall structure of the PID control system depicted in the diagram demonstrates the fundamental principles of feedback control, where the system continuously monitors the process output, compares it to the desired set-point, and adjusts the control signal accordingly to maintain the desired performance. This feedback loop allows the system to adapt to changes in the process, external disturbances, and other factors, ensuring stable and efficient operation.

As depicted in the Fig. 3 and previously mentioned, the infrastructure layer comprises vehicles, base stations, OpenFlow switches, and servers. Multimedia requests originate from vehicles and are directed to servers, while multimedia content is transmitted back from servers to vehicles. Considering the substantial volume of requests, it is crucial to balance server load, prevent overloading, and minimize energy consumption. This ensures that vehicular users receive high-quality multimedia experiences. Servers, also referred to as Physical Machines (PMs), host multiple Virtual Machines (VMs). These VMs are implemented based on NFV technology. NFV will utilize virtualization technologies to transform network functions and services (such as multimedia) into Virtualized Network Functions (VNFs). These VNFs will be implemented in software and executed as VMs on both commodity hardware and high-performance PMs, collectively known as Network Function Virtualization Infrastructures (NFVIs). A VNF may consist of one or more VMs that perform specific tasks on PMs to provide network functions such as multimedia servers. This eliminates the need to purchase and install specialized hardware for network functions. NFV decouples network functions from specialized hardware, allowing these functions to operate as software. VNF instances can be created and deployed dynamically on demand, or they can be dynamically migrated. Multiple instances of VNFs can be scaled up or down based on network conditions. Notably, VM live migration technology enables the remapping of VNFs to different PMs while services remain operational, ensuring no service interruption. The details and architecture of the proposed modular controller's control plane are illustrated in the Fig. 3. The SDN controller for infrastructure management is logically composed of two controllers: a flow controller and a VNF controller, each equipped with a PID controller. The main objective is the simultaneous management of "load" and "energy", balancing the trade-off between load distribution and energy efficiency by consolidating VNFs on fewer PMs when possible. The flow controller is responsible for selecting the best PM from a load perspective to serve multimedia requests, using algorithms like least connection and least response time. To do this, it collects network statistics using OpenFlow messages, then predicts future load and adjusts the load according to a target. The decisions made are sent to the infrastructure for execution using OpenFlow messages. With this approach, the load is intelligently distributed among PMs. Simultaneously, the VNF controller, which is responsible for managing VNFs, complements load distribution by placing VNFs on PMs to optimize resource and energy constraints. To do this, it monitors the resource consumption of PMs using NFV management tools, predicts future resource needs, and calculates the system temperature. A high temperature indicates a shortage of resources and overload. A low temperature indicates an excess of resources and energy waste. The VNF controller reduces overload and energy waste by turning PMs on or off and moving VNFs based on system temperature, using NFV orchestration mechanisms. This is proactive in nature because the controller uses a predictor and minimizes service disruption.

---

[1]The convergence of software-defined networking (SDN) and fifth/sixth generation cellular networks (5G/6G) represents a powerful synergy that is revolutionizing the telecommunications industry. SDN's flexible, programmable, and centralized control of network resources aligns perfectly with the diverse service requirements and performance demands of 5G and emerging 6G networks. By integrating SDN principles, 5G and 6G networks can achieve enhanced network programmability, efficient resource utilization, rapid service deployment, advanced network slicing capabilities, and seamless network function virtualization. This convergence enables network operators to create highly adaptable cellular architectures that can dynamically adapt to evolving user needs and emerging use cases, positioning the industry for continued innovation and progress. The SDN-5G/6G partnership is poised to shape the future of cellular connectivity and redefine the technological landscape.

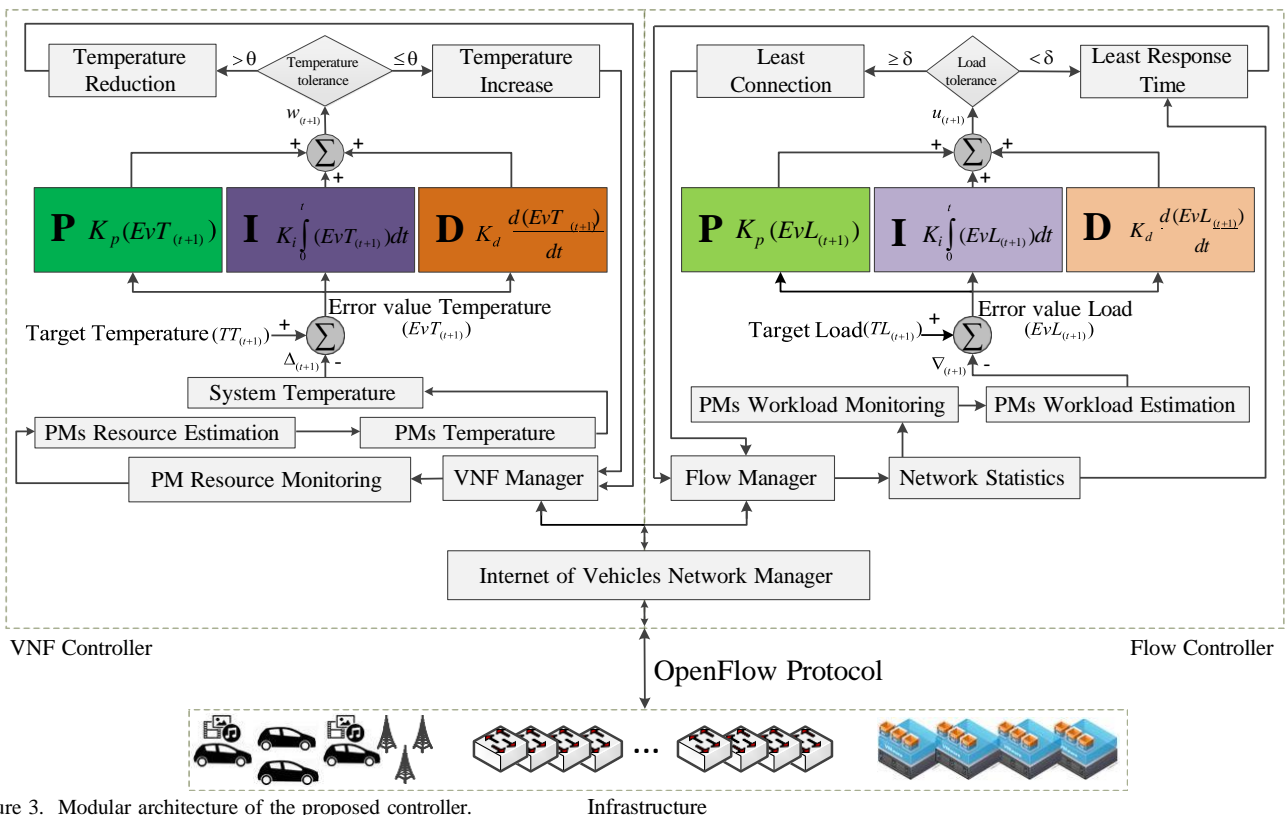

Figure 3. Modular architecture of the proposed controller.

*A. Flow Controller*

This controller performs flow regulation and load balancing operations, generating output signals that achieve equitable workload distribution across PMs. The system initially gathers network performance metrics while continuously tracking and archiving each PM's workload characteristics. These historical records enable predictive modeling of the system's future workload state (at time $t+1$). Now, with a target load in mind, the system's load is adjusted to this desired load. If the system load tolerance exceeds a threshold, then the *Least Connection* algorithm is executed and the server with the fewest connections is selected to send the load to in the next step. This is done to balance the number of connections between PMs. However, if the system load tolerance is below the specified threshold, then the *Least Response Time* algorithm is executed to balance the load between the servers, and the PM with the shortest response time is selected for the next destination. This mechanism maintains load balancing between PMs in most situations and conditions. A general overview of the flow controller algorithm is provided in the Alg. 1, and the details of each of its modules are explained in the following subsections.

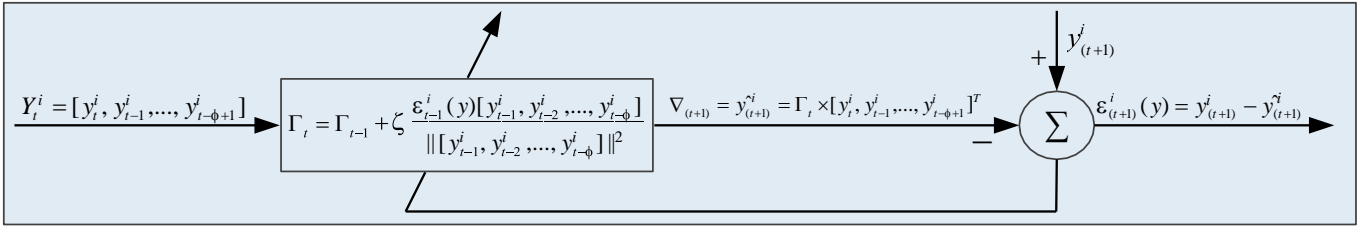

Figure 4. PMs Workload Estimation.

*1) Network Statistics:*

This module periodically collects network statistics from the infrastructure level using OpenFlow messages, such as information about media flows exchanged in the network, their source, destination, path, and network map. Additionally, it gathers PM response times and connection counts using monitoring tools. The strength of the designed controller is to have a comprehensive global view of the events and statistics of the infrastructure level. More accurate decisions are made thanks to the up-to-date and complete statistics collected by this module. One of the most critical statistics continuously monitored by this module is the workload of PMs, obtained via monitoring tools. The module maintains a history of past workload values for each PM, storing up to $\phi$ entries in an array denoted by $Y_t^i = [y_t^i, y_{(t-1)}^i, \ldots, y_{(t-\phi+1)}^i]$. The $y_t^i$ represents the $i$-th PM workload at time $t$.

*2) PMs Workload Estimation:*

Given the array of $Y_t^i$, this module looks for an accurate estimate of the value of $y$ for the next round i.e. $t+1$, which is represented by $\hat{y}_{(t+1)}^i$. For this, we have used a customized *Normalized Least Mean Square (NLMS)* based predictor as shown Fig. 4. The input of this predictor is the array $Y_t^i$ and its output is $\hat{y}_{(t+1)}^i$ (called $\nabla_{(t+1)}$). This closed predictor seeks to reduce the error between the measured value and the estimated value of $y$ ($\epsilon_{(t+1)}^i(y) = y_{(t+1)}^i - \hat{y}_{(t+1)}^i$). The $\hat{y}_{(t+1)}^i$ is obtained by multiplying the filter of $\Gamma_t$ in $[y_t^i, y_{(t-1)}^i, \ldots, y_{(t-\phi+1)}^i]^T$. The mentioned filter is also calculated through the recursive formula $\Gamma_t = \Gamma_{t-1} + \zeta(\epsilon_{t-1}^i(y)[y_{t-1}^i, y_{t-2}^i, \ldots, y_{t-\phi}^i])/||[y_{t-1}^i, y_{t-2}^i, \ldots, y_{t-\phi}^i]||^2$. It should be noted that $\zeta$ is the step size ($0 < \zeta < 2$).

---

**Algorithm 1:** Flow Controller
**Result:** Select the most appropriate PM;
**Data:** – $TL_{(t+1)}$ denotes the desired load level;
 – $\nabla_{(t+1)}$ the estimated system load for time $t+1$;
 – $EvL_{(t+1)}$ the difference between the estimated load level and the target one;
 – $u_{(t+1)}$ is the increment or decrement in the actual load threshold needed to achieve the target utilization;

1 **while** *(as long as the traffic is flowing)* **do**
2    – The *Flow Manager* module receives the incoming packets like Packet-In message;
3    – The *Network Statistics* module extracts the necessary information such as time from the messages;
4    – Response time is given to the *Least Response Time* module;
5    – The *PMs Workload Monitoring* module monitors and records resource consumption of PMs;
6    – The *PMs Workload Estimation* module predicts the system's future workload ($\nabla_{(t+1)}$);
7    – The *PID Controller* module calculates $u_{(t+1)}$ with regard to $\nabla_{(t+1)}$, $TL_{(t+1)}$, and $EvL_{(t+1)}$;
8    **if** $(u_{(t+1)} \geq \delta)$ **then**
9      – *Least Connection* method runs;
10      – The PM with the least load is selected;
11    **else**
12      – *Least Response Time* method runs;
13      – The PM with the least response time is selected;
14    **end**
15 **end**

---

*3) PID Controller for load:*

In the SDN controller, we use the PID controller to decide how to balance the load between PMs. Given the predicted load ($\nabla_{(t+1)}$), the PID controller seeks to fine-tune the load to the target load ($TL_{(t+1)}$). This work introduces a control theory approach that dynamically adjusts the load based on system behavior and target utilization levels. This enables the SDN controller to adapt and learn from ongoing system performance. We assess a PID controller that modifies the load threshold, thereby altering the load level the system can handle, to achieve the desired resource utilization for each dimension (e.g., CPU) more safely. Here, $TL_{(t+1)}$ represents the target load level, $\nabla_{(t+1)}$ the estimated level, $EvL_{(t+1)}$ the discrepancy between the current load and the target load, and $u_{(t+1)}$ the adjustment in the load threshold necessary to meet the target utilization. The value of $u_{(t+1)}$ at time $t + 1$ is determined by Equation 1:

$$u_{(t+1)} = K_p(EvL_{(t+1)}) + K_i \int_0^t (EvL_{(t+1)})dt + K_d \frac{d(EvL_{(t+1)})}{dt} \quad (1)$$

In this equation, increasing the proportional component ($K_p$) results in a faster response but can also cause overshooting and oscillation issues. Enhancing the integral component ($K_i$) minimizes steady-state errors but can lead to increased oscillations. The derivative component ($K_d$) helps to reduce oscillations, although it may result in a slower response. Simply adjusting these values proved insufficient for achieving optimal performance, characterized by a stable and swift final response. The filter constrains threshold load levels to a range between 0.2 and 0.8, rather than the full [0, 1] range. By limiting the range of feasible values for the load threshold, the filter reduces fluctuations due to rapid switching between accepting too many and too few services.

If the tolerance of the load of PMs is less than the threshold ($\delta$), the *Least Response Time* algorithm is executed, and if it is greater than the mentioned threshold, the *Least Connection* algorithm is executed.

The *Least Response Time* approach utilizes data provided by the *Network Statistics* module to identify the PM that exhibits the lowest response latency at a given moment. In contrast, the *Least Connection* method directs incoming requests to the PM currently handling the fewest active sessions. Consequently, in scenarios where load disparity among PMs is pronounced, the *Least Connection* approach excels at swiftly pinpointing underutilized resources and distributing workload accordingly. Conversely, when load equilibrium is paramount, the *Least Response Time* strategy proves more effective in rapidly stabilizing system performance.

*4) Least Response Time:*

As mentioned earlier, this module selects the PM with the least response time according to the Alg. 2 (where $n$ is the number of PMs).

*5) Least Connection:*

This module determines the PM with the fewest active connections as the target for incoming multimedia requests, based on the algorithm described below.

It is important to note that at the initial time (time zero), the first destination is chosen randomly, while subsequent

---

**Algorithm 2:** Least Response Time

1  The response time for PM 0 is assumed to be zero,
   **for** *(i = 1 to n, i + +)* **do**
2     **if** *(the response time for PM i is less than the response time for PM i − 1)* **then**
3        {PM *i* is the destination PM};
4     **else**
5        {PM *i* − 1 is the destination};
6     **end**
7  **end**

---

**Algorithm 3:** Least Connection

1  The number of connections for PM 0 is assumed to be zero,
2  **for** *(i = 1 to n, i + +)* **do**
3     **if** *(the number of connections for PM i is less than the number of connections for PM i − 1)* **then**
4        {PM *i* is the destination PM};
5     **else**
6        {PM *i* − 1 is the destination};
7     **end**
8  **end**

---

destinations are selected in accordance with Alg. 3.

Finally, the above decisions are translated by the *Flow Manager* module into OpenFlow commands and are handed over to the infrastructure layer for execution.

### B. VNF Controller

Concurrently, the VNF controller manages both the PMs and the optimal placement of existing multimedia VNFs within the system, pursuing two primary objectives: avoiding system overload and minimizing energy consumption. To achieve this, it continuously monitors the resource utilization of the PMs[2] and predicts their future usage. Each PM's consumption level is mapped to a corresponding temperature value, which is then used to compute the overall system temperature (aggregated across all PMs). The controller strives to regulate this system temperature toward a predefined target. In overload conditions, where resource utilization is high, the system temperature rises, prompting the execution of the *temperature reduction algorithm*. Conversely, during underload conditions—characterized by low resource utilization and energy inefficiency—the *temperature increase algorithm* is applied. Through this temperature regulation mechanism, the system effectively mitigates both overload and unnecessary energy consumption. The general procedure is outlined in Alg. 4, with further details provided in the following subsections.

*1) PM Resource Monitoring:*

This module samples the CPU utilization of the active PMs at specific time intervals and maintains $\varphi$ of the most recent sampled values. These values are stored in the array $X^i_t = [x^i_t, x^i_{t-1}, ..., X^i_{t-\varphi+1}]$. The value $x^i_t$ represents the PM $i$'s sampled value at time $t$.

*2) PMs Resource Estimation:*

This module seeks to estimate the future CPU utilization of each active PM ($\hat{x}^i_{(t+1)}$). Following Fig. 5, this is done based on the historical CPU utilization records that have been maintained, as well as the NLMS predictor. The input to this closed-loop predictor is $X^i = [x^i_t, x^i_{t-1}, ..., X^i_{t-\varphi+1}]$, and the output is $\hat{x}^i_{(t+1)}$, which aims to minimize the error ($\varepsilon^i_{(t+1)}(x) = x^i_{(t+1)} - \hat{x}^i_{(t+1)}$). The value of $\hat{x}^i_{(t+1)}$ is obtained by multiplying the filter $Y_t$ with $[x^i_t, x^i_{t-1}, ..., x^i_{t-\varphi+1}]^T$. This filter is obtained recursively through the relationship

---

[2]This study focuses solely on CPU usage; however, future research may incorporate additional PM resources.

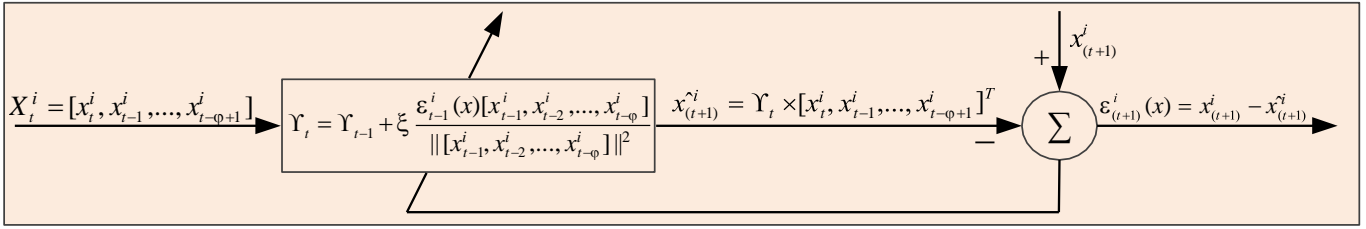

Figure 5. PMs Resource Estimation.

$\Upsilon_{t-1} + \xi \frac{\varepsilon_{t-1}^i(x)[x_{t-1}^i, x_{t-2}^i, ..., x_{t-\varphi}^i]}{\|[x_{t-1}^i, x_{t-2}^i, ..., x_{t-\varphi}^i]\|^2}$, where $\xi$ is the step size, which varies between 0 and 2.

*3) PMs Temperature:*

The temperature of each active PM is affected by the $\hat{x}_{(t+1)}^i$. If $\hat{x}_{(t+1)}^i$ is less than the threshold $\alpha$, then we consider $Temp_{(t+1)}^i$ (or the temperature of PM $i$) as 0. If $\hat{x}_{(t+1)}^i$ is greater than the threshold $\beta$, then we consider $Temp_{(t+1)}^i$ as 1. Finally, if $\hat{x}_{(t+1)}^i$ is between the two mentioned thresholds, then $Temp_{(t+1)}^i = \frac{\hat{x}_{(t+1)}^i - \alpha}{\beta - \alpha}$. This approach leads to a smoothing of the PM temperatures and also ensures that the temperature is proportionate to the CPU utilization of the PM.

*4) System Temperature:*

The overall system temperature ($\Delta_{(t+1)}$) is equal to the average temperature of the active PMs, or in other words $\Delta_{(t+1)} = \frac{\sum_{i=1}^{n} Temp_{t+1}^i}{n}$ where $n$ is the total number of active PMs in the system. In this regard, the PID controller aims to regulate the system temperature around the Target Temperature ($TT_{(t+1)}$).

*5) PID Controller for Temperature:*

In this study, we introduce a control theory approach that adaptively adjusts the system temperature in response to real-time system behavior and target utilization levels. This mechanism enables the SDN controller to incrementally learn and adapt based on the system's operational behavior. We implement and evaluate a PID controller, which dynamically modifies the temperature threshold—i.e., the maximum acceptable system temperature—to ensure resource utilization (specifically CPU) remains within safe and efficient bounds. In this context, $TT_{(t+1)}$ represents the target temperature, $\Delta_{(t+1)}$ denotes the observed temperature, $EvT_{(t+1)}$ is the deviation between the current and target temperatures, and $w_{(t+1)}$ indicates the required adjustment (positive or negative) to the temperature threshold in order to maintain the desired utilization. The value of $w_{(t+1)}$ at time $t+1$ is computed using Equation 2:

$$w_{(t+1)} = K_p(EvT_{(t+1)}) + K_i \int_0^t (EvT_{(t+1)})dt + K_d \frac{d(EvT_{(t+1)})}{dt} \quad (2)$$

In this equation, increasing the $P$ component ($K_p$) leads to faster response but also to overshooting and oscillation problems. Increasing $I$ component ($K_i$) reduces stationary errors but at the expense of larger oscillations. Finally, the $D$ component ($K_d$) reduces the oscillations but may lead to slower response. Only setting these values turned out to be insufficient to achieve a good performance (stable and fast response final value ($P + I + D$)). This filter keeps threshold load levels between certain values, i.e., between 0.2 and 0.8 instead of the [0, 1] range. Limiting the spectrum of feasible values for the temperature threshold reduces fluctuations caused by fast switching between accepting too many and too few services.

Finally, if the temperature tolerance is less than a certain threshold ($\theta$), the *temperature increase* algorithm is executed. On the other hand, if the temperature tolerance is greater than the mentioned threshold, the *temperature decrease* algorithm is executed.

*6) Temperature Increase:*

The objective is to maintain the system temperature—reflecting the load on active PMs—within a normal operating range, avoiding both excessively low and high values. When the system temperature falls below a predefined threshold $\theta$ (i.e., $w_{(t+1)} \leq \theta$), it indicates low CPU utilization across active PMs, leading to unnecessary energy consumption. To address this, some underutilized or idle PMs should be powered down to bring the system temperature back to the desired range. In essence, by deactivating certain low-utilization PMs, energy usage is optimized and the system temperature is elevated from a low state to a normal level. It is important to note that each PM hosts multiple multimedia VNFs, and before shutting down any PM, the VNFs it contains must be migrated to alternative PMs. Alg. 5 outlines the procedure for this migration.

$n$ is the total number of PMs, and the aim is to turn off approximately half of the PMs and migrate their VNFs to the remaining half.

*7) Temperature Reduction:*

In the opposite scenario, where the system temperature surpasses the threshold $\theta$ (i.e., $w_{(t+1)} > \theta$), it reflects high resource utilization across the PMs and an increased risk of overload due to potential resource saturation. To mitigate this, the optimal strategy is to activate additional PMs, thereby distributing the workload more evenly and lowering the system temperature from an elevated state to a normal range. This process is governed by Alg. 6.

It is important to note that using this method results in an increase of $n/2$ in the number of PMs, with half of the VNFs from an existing active PM being migrated to the new PM.

Ultimately, the decisions made in this process are translated by the *VNF Manager* module into OpenFlow commands and are sent to the infrastructure layer for execution.

**Algorithm 4:** VNF Controller

**Result:** Precise adjustment of the system temperature to prevent overload and energy wastage in the PMs;

**Data:**
– $TT_{(t+1)}$ denotes the desired temperature level;
– $\Delta_{(t+1)}$ the estimated system temperature for time $t + 1$;
– $EvT_{(t+1)}$ the difference between the estimated temperature level and the target one;
– $w_{(t+1)}$ is the increment or decrement in the actual temperature threshold needed to achieve the target utilization.

1 **while** *(as long as the traffic is flowing)* **do**
2   – The *VNF Manager* module receives the incoming packets like Packet-In message; It also collects VNF information and manages them according to the received commands;
3   – The *PM Resource Monitoring* module monitors and records the resource consumption of the PMs (specifically the CPU usage);
4   – The *PMs Resource Estimation* module predicts the resource consumption of the PMs (CPU usage) for the next round, specifically at time $t + 1$, using the NLSM algorithm;
5   – The *PMs Temperature* module considers the corresponding temperature for each PM based on its resource consumption;
6   – The *System Temperature* module considers the overall temperature of the system, including all PMs ($\Delta_{(t+1)}$), which reflects the overall state of the system: underload (low temperature), normal load (desired temperature), and overload (high temperature);
7   – The *PID Controller* module calculates $w_{(t+1)}$ with regard to $\Delta_{(t+1)}$, $TT_{(t+1)}$, and $EvT_{(t+1)}$;
8   **if** $(w_{(t+1)} \geq \theta)$ *(i.e., high system temperature)* **then**
9     – *Temperature Reduction* method runs;
10    – The number of PMs is increased, resulting in a decrease in system temperature and the prevention of overload;
11  **else**
12    – *Temperature Increase* method runs;
13    – The number of PMs is reduced, resulting in an increase in system temperature and preventing energy wastage;
14  **end**
15 **end**

**Algorithm 5:** Temperature Increase

1 **for** $(i = 1$ *to* $n/2, i + +)$ **do**
2   **for** $(j = n$ *to* $n/2, j - -)$ **do**
3     **while** *(until PM i still has VNFs and $i \neq j$)* **do**
4       {Migrating a VNF from PM $i$ to PM $j$ (where $j \neq i$)};
5     **end**
6     {PM $i$ should be powered off};
7   **end**
8 **end**

**Algorithm 6:** Temperature Reduction

1 **for** $(i = 1$ *to* $n/2, i + +)$ **do**
2   A new PM should be powered on;
3   **while** *(until the number of VNFs on PM i is reduced to half)* **do**
4     {Migrating a VNF from PM $i$ to the new PM};
5   **end**
6 **end**

## IV. IMPLEMENTING AND PERFORMANCE EVALUATION

### A. Experimental setup

The proposed IoV network plan, illustrated in Fig. 6, is derived from the framework presented earlier. The network's core component, the SDN controller, can be implemented using various open-source frameworks such as RYU, NOX, or POX. This controller leverages the OpenFlow protocol to establish a centralized control plane, enabling dynamic network configuration and optimization. The data plane, comprising multimedia servers and network switches, can be simulated using Mininet[3] or realized using OpenStack. These components are responsible for data transmission, processing, and storage within the network. The infrastructure layer, designed to emulate real-world conditions, is implemented using the SUMO traffic simulator. This simulation tool enables the generation of realistic traffic scenarios, incorporating heterogeneous vehicle types with different communication capabilities and mobility patterns. By modeling a range of traffic conditions—including peak periods, non-peak hours, and emergency situations—researchers can systematically assess the network's performance and resilience across diverse operational contexts.

The actual implementation of the proposed IoV network plan is depicted in Fig. 7, utilizing the G9 and G10 servers and systems. This real-world testbed allows for rigorous evaluation

---
[3]Includes User Plane Function (UPF) emulation for N6 traffic.

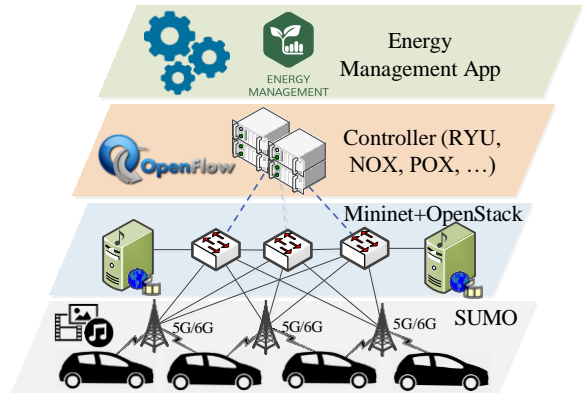

Figure 6. IoV network plan: smart vehicles, RSUs, switches, controller, multimedia servers, and multimedia streams.

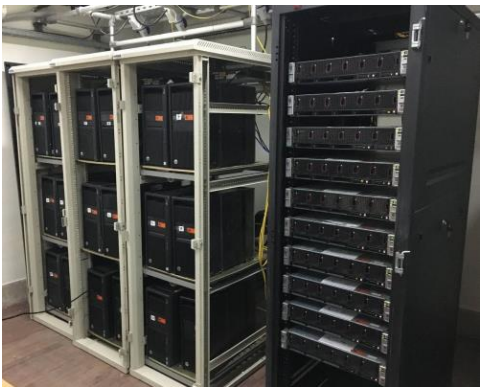

Figure 7. Testbed including multimedia servers, controller, smart virtual vehicles, and OpenFlow switches.

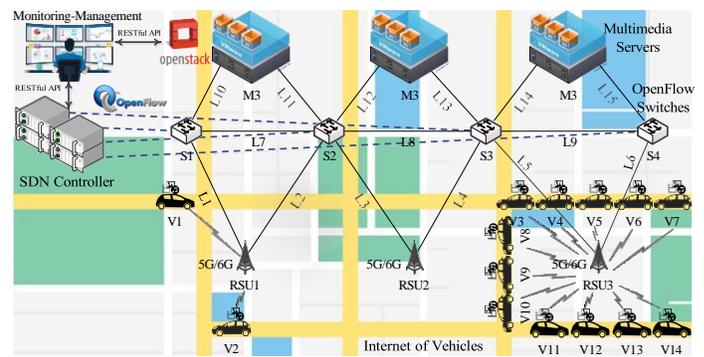

Figure 9. Traffic scenarios with realistic urban settings with multiple intersections and road segments (snapshot of SUMO).

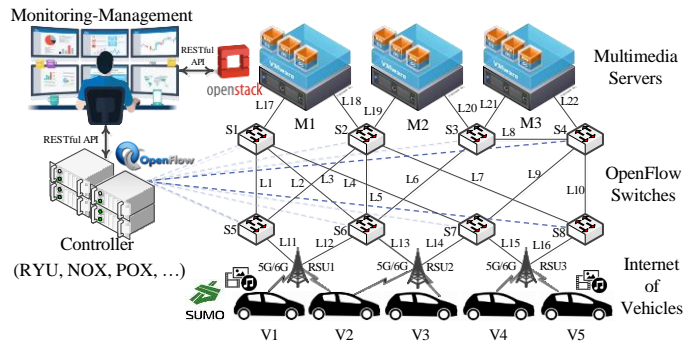

Figure 8. IoV network topology, communications, components and links (5GC abstracted).

and validation of the network's performance and scalability under realistic conditions.

The detailed architecture and interconnection of the proposed network components are depicted in Fig. 8. Three high-performance hardware-based multimedia servers (M1, M2, and M3) are utilized to host a large number of multimedia VNFs. The virtualization layer is facilitated by OpenStack, which allows for the dynamic deployment and efficient management of these VNFs. Communication between vehicles and servers is facilitated by three Roadside Units (RSUs 1, 2, and 3)[4], eight OpenFlow switches (S1 to S8), and a network of twenty-two links (L1 to L22). This infrastructure enables the establishment of multimedia flows between various network components. The urban environment and vehicle traffic are simulated using the SUMO traffic simulator. This allows for the realistic modeling of real-world traffic conditions, including varying traffic densities, diverse vehicle behaviors, and complex road network topologies. The OpenFlow switches are controlled by a central SDN controller, which enables dynamic network configuration and optimization. The controller leverages the OpenFlow protocol to establish a centralized control plane, allowing efficient resource allocation and traffic management. Additionally, a RESTful API is used to monitor and manage the servers and controller, providing a centralized view of the network's overall health and performance. This API enables remote configuration, troubleshooting, and performance analysis. Fig. 9 displays a snapshot from the SUMO-generated traffic simulation, offering a visual depiction of prevailing traffic conditions. This visualization facilitates the analysis of real-time network dynamics, including vehicle mobility, traffic flow patterns, and the influence of different traffic scenarios on the overall performance of the IoV system.

### B. Results

This section presents the results and their subsequent analysis and interpretation. The results are divided into two subsections: data plane and control plane.

*1) Data plane assessment:* In this subsection, we present a comprehensive analysis of servers (M1-M3) and switches (S1-S8) performance under two operational modes: *traditional* and *proposed*. To assess performance under various traffic conditions, five scenarios were defined, ranging from *very low* to *very high* load, based on the number of vehicles and traffic volume. The results of this analysis are visualized in Fig. 10 and tabulated in Table I. Fig. 10a illustrates that all three PMs' CPU utilization remains balanced across all scenarios in the proposed mechanism, indicating effective load balancing, even under heavy traffic conditions. This contrasts with the traditional method, which lacks SDN and NFV-based prediction and management, leading to uneven load distribution. It is noteworthy that the proposed controller's simultaneous management of resources and energy results in an adaptive deployment strategy. In scenarios with low to moderate traffic, only a subset of PMs (M1 and M2) are activated. Only under the most demanding, very high-traffic scenario, are all three PMs (M1, M2, and M3) deployed. This dynamic resource allocation demonstrates the efficiency and adaptability of the proposed approach. Also, this adaptive resource allocation strategy contributes to achieving high throughput by the active PMs (Fig. 10b). Fig. 10c demonstrates the effectiveness of the proposed PID controller in regulating temperature and, consequently, power consumption of the PMs. Across all scenarios, the proposed mode consistently exhibits lower power consumption compared to the traditional approach, indicating superior energy efficiency.

Statistical analyses of delay, setup time, and jitter for PM connections under various conditions are illustrated in Figs. 10d through 10f. The proposed model consistently outperforms the conventional method, demonstrating significantly lower values across all three metrics. This enhancement in network performance directly contributes to improved Quality of Service (QoS) for vehicle users, particularly regarding multimedia streaming. The Quality of Experience (QoE) evaluation metrics—Mean Opinion Score (MOS), Rating factor (R-factor), and multimedia loss rate—are presented in Figs. 10g to 10i. Higher MOS (ranging from 1 to 5) and R-factor (ranging from 1 to 100) values correspond to superior quality and user experience. The proposed framework achieves more robust QoE results, with higher MOS and R-factor scores and reduced multimedia loss rates compared to the traditional approach. This improvement in QoE is attributed to the proactive and predictive characteristics of the proposed mechanism. Specifically, the average MOS for the proposed framework is 4.6, whereas the traditional approach achieves approximately 3.5, despite the fact that even in the worst-case scenario, the multimedia loss rate in the proposed method remains below 10%.

---
[4]RSUs interface with the 5G core network (5GC) via N2/N3, with UPF aggregating mobility-opaque IP flows over N6 to the SDN data network [11].

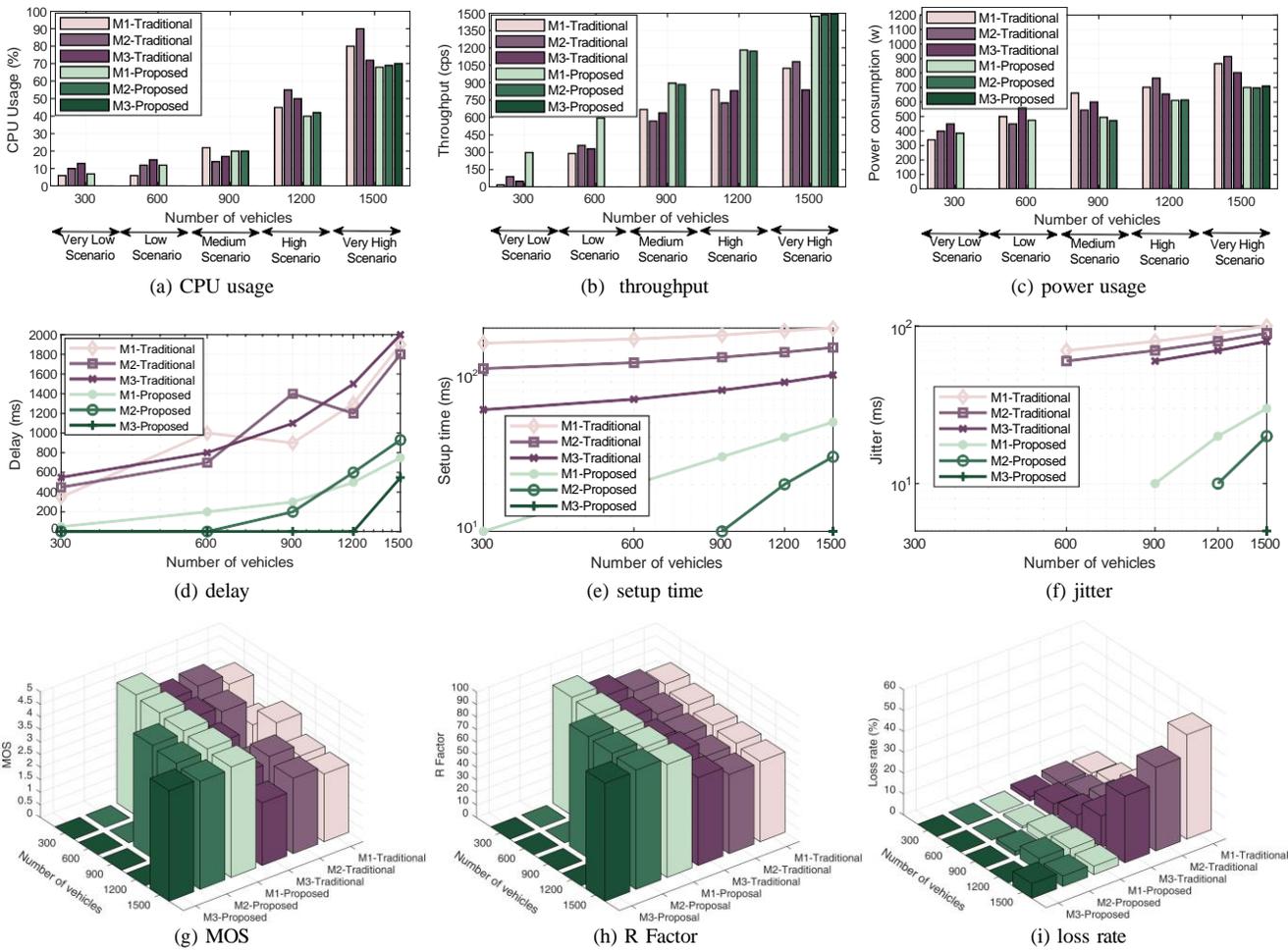

Figure 10. Comparative analysis of resource utilization, efficiency, energy consumption, load balancing, QoS (such as delay, jitter), and QoE (such as MOS, R Factor) parameters across different methods and traffic conditions.

The success of the proposed approach can be primarily attributed to its effective management of VNFs based on traffic patterns. This relationship is clearly illustrated in Fig. 11. Specifically, when the number of vehicles—and consequently the traffic volume and the number of switches—is low, the number of active VNFs is also correspondingly reduced.

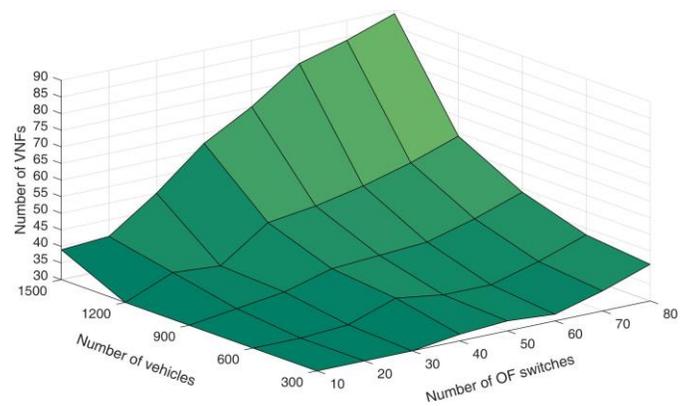

Figure 11. The pattern of increasing the number of VNFs corresponds to the increase in the number of vehicles and OF switches.

Conversely, as these parameters increase, the number of VNFs adjusts dynamically to accommodate the demand. This adaptive mechanism enables the network to scale efficiently while minimizing energy consumption.

In addition to examining the data plane PMs, we also analyze another key data plane component: the switches (S1–S8),

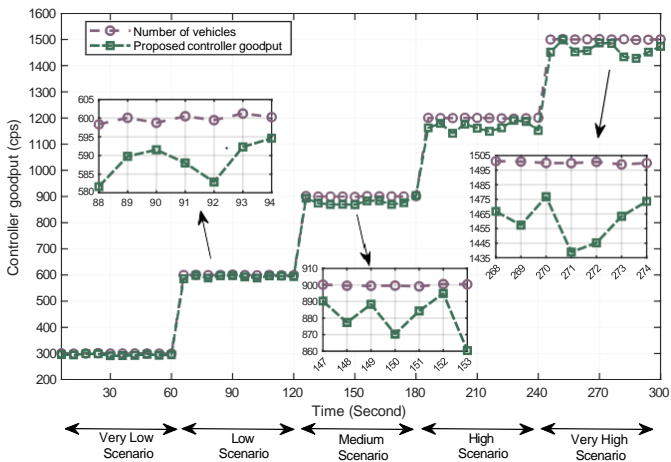

Figure 12. Scalability of the proposed controller.

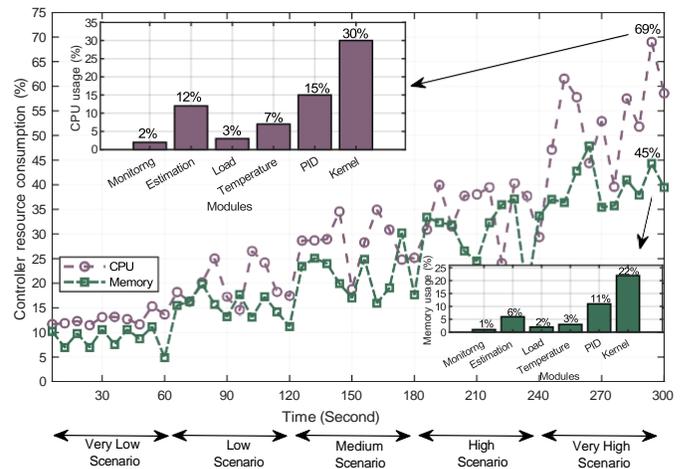

Figure 13. Efficiency of the proposed controller.

as outlined in Table I. The table reports the CPU utilization levels of these network switches. Under traffic management by the SDN controller—represented by the green columns—the CPU usage across switches is both lower and more balanced compared to the traditional approach, indicated by the pink columns. For example, in a very high traffic scenario using the traditional method, switch S1 exhibits a CPU utilization of 70.8%, whereas switch S2 shows only 19.8%, revealing a pronounced imbalance in resource allocation. Conversely, in the same high-traffic context, the proposed controller's centralized management yields a more equitable distribution, with CPU utilizations for S1 and S2 closely aligned at 40.5% and 40.8%, respectively, indicating a well-balanced workload across the switches.

*2) Control plane assessment:* In this subsection, we evaluate the scalability and efficiency of the designed controller. The results are illustrated in Figs. 12 and 13, respectively. Traffic is generated for each vehicle, increasing incrementally based on the scenario pattern as the number of vehicles grows. The controller's goodput closely aligns with the traffic volume, demonstrating its ability to handle nearly all vehicle requests effectively. For instance, in the $90^{th}$ second, the controller successfully responded to 592 out of 599 requests. Similarly, under heavy traffic conditions in the $274^{th}$ second, it managed 1475 out of 1500 requests. These results highlight the controller's scalability and its capacity to maintain high performance even under significant traffic loads. The CPU and memory utilization of the proposed controller are illustrated in Fig. 13. The average increase in resource consumption follows a linear trend closely aligned with traffic growth. Importantly, even under peak load conditions, the controller's CPU and memory usage remain below 69% and 45%, respectively, demonstrating that it does not constitute a performance bottleneck. Detailed values are shown in the corresponding subfigures. Analysis indicates that the operating system kernel is responsible for the majority of resource consumption—approximately 30% of CPU and 22% of memory—while the custom-designed modules consume considerably less. Among these modules, the PID controller has the highest resource usage, accounting for 15% CPU and 11% memory utilization. In contrast, the monitoring module demands the least resources, using only 2% CPU and 1% memory. The estimation module, which implements the NLMS algorithm, exhibits moderate resource requirements. Notably, none of the designed modules consume more than half the resources attributed to the kernel, indicating that they impose minimal additional overhead on the controller.

## V. Conclusion and Future Directions

This study introduced an innovative energy-efficient framework for multimedia resource management in software-defined 5G/6G IoV networks, leveraging a PID controller-based approach. Through comprehensive performance evaluations, the proposed framework demonstrated its capability to optimize load balancing, reduce energy consumption, and ensure high-quality multimedia experiences for vehicular users. By integrating SDN and NFV technologies, the framework effectively addressed the scalability and adaptability challenges posed by the exponential growth of multimedia traffic in IoV networks. The key findings emphasize the proposed framework's advantages compared to conventional methods. The adaptive load distribution mechanism not only improved server utilization and reduced power consumption but also ensured near-optimal CPU usage across network switches. Moreover, the dynamic, centralized temperature management achieved by the PID controller ensured balanced resource allocation, significantly reducing energy waste, and sustained QoS even under heavy traffic conditions. The framework's predictive capabilities, combined with efficient VNF management, underline its robustness and practical applicability for next-generation IoV systems. Building on the promising outcomes of this research, several avenues for future exploration are identified:

– *Extension to Multi-Access Edge Computing (MEC)*: Future work could explore integrating MEC technologies with the proposed framework to further enhance latency-sensitive multimedia applications and enable localized processing closer to end-users.

– *Incorporation of AI and Machine Learning (ML)*: Although the PID controller provides efficient management, integrating AI/ML algorithms has the potential to improve

Table I
ANALYZING THE CPU USAGE OF SWITCHES UNDER DIFFERENT TRAFFIC CONDITIONS AND METHODS

| Method: → | | | Traditional | | | | | | | | Proposed | | | | | | | |
|---|---|---|---|---|---|---|---|---|---|---|---|---|---|---|---|---|---|---|
| Scenario: ↓ | | Switch | S1 | S2 | S3 | S4 | S5 | S6 | S7 | S8 | S1 | S2 | S3 | S4 | S5 | S6 | S7 | S8 |
| Number of vehicles: → | Very low | 300 | 3.1 | 13.2 | 8.6 | 5.5 | 15.4 | 2.2 | 7.6 | 14.7 | 3.6 | 3.1 | 4.6 | 3.6 | 4.8 | 3.4 | 3.7 | 4.9 |
| | Low | 600 | 8.8 | 19.8 | 25.8 | 10.4 | 15.7 | 22.3 | 11.7 | 13.7 | 10.8 | 11.7 | 11.4 | 10.4 | 11.5 | 10.7 | 11.4 | 10.7 |
| | Medium | 900 | 23.7 | 31.7 | 10.5 | 8.8 | 28.7 | 20.9 | 10.5 | 19.9 | 20.3 | 21.6 | 21.4 | 20.5 | 21.7 | 21.7 | 20.9 | 20.7 |
| | High | 1200 | 10.7 | 41.6 | 35.6 | 30.5 | 11.6 | 40.4 | 26.7 | 33.6 | 30.4 | 32.7 | 31.9 | 30.5 | 30.5 | 30.8 | 30.7 | 31.7 |
| | Very high | 1500 | 70.8 | 19.8 | 25.7 | 41.7 | 51.5 | 60.3 | 35.2 | 50.7 | 40.5 | 40.8 | 40.0 | 41.7 | 42.6 | 41.0 | 40.6 | 40.8 |

predictive precision and facilitate intelligent decision-making in dynamic network scenarios.

– *Comprehensive Resource Optimization*: Broadening resource monitoring to encompass memory, storage, and network bandwidth would enable a more comprehensive optimization strategy for IoV systems.

– *Support for Diverse Traffic Patterns*: Examining the framework's adaptability to evolving traffic conditions—such as environments with a mix of autonomous and human-driven vehicles—can help ensure its robustness across a variety of real-world applications.

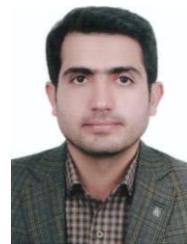

**Ahmadreza Montazerolghaem** is an Assistant Professor in the Department of Information Technology, Faculty of Computer Engineering, at the University of Isfahan. He holds a Ph.D. in Computer Engineering, graduated with distinction from Ferdowsi University of Mashhad. His research focuses on SDN/NFV, IoT/IoV, 5G/6G, and multimedia networking. Recognized as among the top 2% of scientists worldwide in 2025 (Stanford/Elsevier), he has authored numerous high-impact publications in leading journals, including the IEEE Internet of Things Journal, where he also serves as an Associate Editor.